\documentclass[extra,onecolumn]{gji}
\usepackage[fleqn]{amsmath}
\usepackage{amsfonts,amssymb}
\usepackage{timet,epsfig,wrapfig,psfrag}
\usepackage{natbib}
\usepackage{symbol}
\usepackage{color}
\usepackage[normalem]{ulem}
\usepackage{booktabs}
\usepackage{multirow} 
\usepackage{pdfcolmk} 
\usepackage{hyperref}
\hypersetup{
    colorlinks=true,
    linkcolor=blue,
    filecolor=magenta,      
    urlcolor=blue,
    }

\newcommand{\w}{\mathbf}

\begin{document}

\bibliographystyle{chicago}


\title[Introduction to phase transmission fibre-optic sensing of seismic waves]
{Introduction to phase transmission fibre-optic sensing of seismic waves}

\author[Andreas Fichtner et al.]{
\parbox{\linewidth}{Andreas Fichtner$^1$, Adonis Bogris$^2$,  Thomas Nikas$^3$,  Daniel Bowden$^1$, Konstantinos Lentas$^4$, Nikolaos S.  Melis$^4$, Christos Simos$^5$, Iraklis Simos$^6$ and Krystyna Smolinski$^1$} \\
$^1$ Department of Earth Sciences, ETH Zurich, Switzerland\\
$^2$ Department of Informatics and Computer Engineering, University of West Attica, Egaleo, Greece\\
$^3$ Department of Informatics and Telecommunications, National and Kapodistrian University of Athens, Athens, Greece\\
$^4$ National Observatory of Athens, Institute of Geodynamics, Greece\\
$^5$ Department of Physics, Electronics and Photonics Laboratory,  University of Thessaly, Lamia, Greece\\
$^6$ Department of Electrical and Electronics Engineering, University of West Attica, Egaleo, Greece}

\maketitle


\begin{summary}
This manuscript is concerned with phase changes of signals transmitted through deforming optical fibres. As a first result, it establishes an exact relation between observable phase changes and the deformation tensor along the fibre. This relation is non-linear, and it includes effects related to both local changes in fibre length and deformation-induced changes of the local speed of light or refractive index.\\[5pt]
In seismic applications, where the norm of the earthquake-induced deformation tensor is orders of magnitude smaller than $1$, a useful first-order relation can be derived. It simply connects phase changes to an integral over in-line strain along the fibre times the local refractive index. Under the assumption that spatial variations of the refractive index are fast compared to the seismic wavelength, this permits a direct synthesis of phase change measurements from distributed strain measurements, for instance, from Distributed Acoustic Sensing (DAS).\\[5pt]
A more detailed analysis of the first-order relation reveals that a perfectly straight fibre would produce zero phase change measurements, unless deformation was sufficiently widespread to affect the starting and end points of the fibre. If this condition is not met, non-zero measurements can only result from curvature of the fibre or from a heterogeneous distribution of the refractive index. Segments of the fibre that are strongly curved generally make larger contributions to the observed phase change.\\[5pt]
Second- or higher-order effects, not captured by the first-order relation, may be observable in specific scenarios, including deformation exactly perpendicular to the fibre orientation. They are associated with a frequency content that is higher than that of the underlying deformation field. Though higher-order effects may be realised in controlled laboratory settings, they are unlikely to occur in seismic experiments where fibre geometries are irregular and waves asymptotically propagate in all directions with all possible polarisations, as a consequence of 3-D heterogeneity \\[5pt]
%
\end{summary}


\section{Introduction}\label{S:Introduction}

Distributed Acoustic Sensing (DAS) is a family of technologies to measure deformation along a fibre-optic cable using interferometry of back-scattered laser light \citep{Hartog_2017}. Following early applications in perimeter security or traffic and pipeline monitoring \citep[e.g.][]{Owen_2012,Hill_2015}, DAS became a widely used tool in seismic exploration and monitoring, where optical cables are often pre-installed in boreholes \citep[e.g.,][]{Mateeva_2013,Mateeva_2014,Daley_2013,Daley_2014,Daley_2016,Li_2015,Dean_2016,Hornmann_2016}. The ability of DAS to record deformation in a broad frequency range from mHz to kHz with dense spatial sampling at metre scale \citep[e.g.,][]{Lindsey_2020,Paitz_2021}, makes it attractive also for seismological applications. The ability to piggy-back on existing fibre-optic telecommunication infrastructure has led to novel applications in urban seismology, with a focus on seismic hazard in densely populated areas, where large numbers of conventional seismic instruments may be difficult or expensive to deploy \citep[e.g.][]{Lindsey_2017,Martin_2017,Biondi_2017,Ajo_2019,Spica_2020,Yang_2021}. The relative ease of deploying fibre-optic cables in challenging terrain, enables seismological studies on glaciers \citep{Walter_2020,Klaasen_2021}, volcanoes \citep{Klaasen_2021,Currenti_2021,Klaasen_2022}, and avalanche-prone slopes \citep{Lindner_2021} that would not have been possible without DAS. In parallel, theoretical developments improved our understanding of how DAS data may be exploited, for instance, by ambient field interferometry \citep{Paitz_2019}.\\[5pt]
While back-scattering allows DAS to achieve distributed measurements with an effective channel spacing in the centimetre range, it also limits the length of the fibre that can be interrogated, typically to a few tens of kilometres. Light intensity loss with increasing propagation distance along the fibre decreases the signal-to-noise ratio. Though emerging amplifier technologies may help to reduce this problem, the installation of amplifiers along existing telecommunication cables or in harsh terrain may not always be possible.\\[5pt]
Emerging alternative systems overcome this limitation by measuring deformation-induced changes in the phase \citep{Marra_2018,Bogris_2021,Bogris_2022,Bowden_2022} or the polarisation \citep{Mecozzi_2021} of transmitted laser light. The ability of transmission-based systems to achieve interrogation distances of hundreds or thousands of kilometres opens new opportunities to investigate seismic activity and Earth structure in remote regions where conventional seismic instrumentation is sparse. This includes, most importantly, the oceans and polar regions. The main drawback of transmission-based systems, however, lies in the averaging of deformation along the fibre. Phase or polarisation changes are accumulated along the fibre, apparently erasing information about the location where the underlying deformation occurred. Hence, in contrast to DAS, the measurement is not distributed but integrated. As a consequence, it may be more challenging to use transmission measurements to infer Earth structure or earthquake locations.\\[5pt]
In the following sections, we develop a theory for the calculation of observed optical phase changes caused by fibre deformation. Section \ref{S:general} sets the general stage and leads to an approximation-free equation that relates the deformation field to phase change measurements. For typical seismic wavefields, where strain is much smaller than $1$, this relation can be linearised, thereby producing various first-order approximations that can be found in section \ref{S:first}. One of these approximations allows us to easily forward model phase change measurements and to synthesise them from DAS data, thereby enabling a comparison of the two measurement systems. Another one highlights the role of cable curvature, showing that the sensitivity of a fibre segment to deformation is proportional to the local curvature. The more a fibre is curved, the better it records deformation. Finally, in section \ref{S:Second}, we investigate under which conditions higher-order effects may be observable, coming to the conclusion that they can safely be ignored in most seismological applications.

\section{General developments}\label{S:general}

\begin{center}
\begin{figure}
\center\scalebox{0.48}{\includegraphics{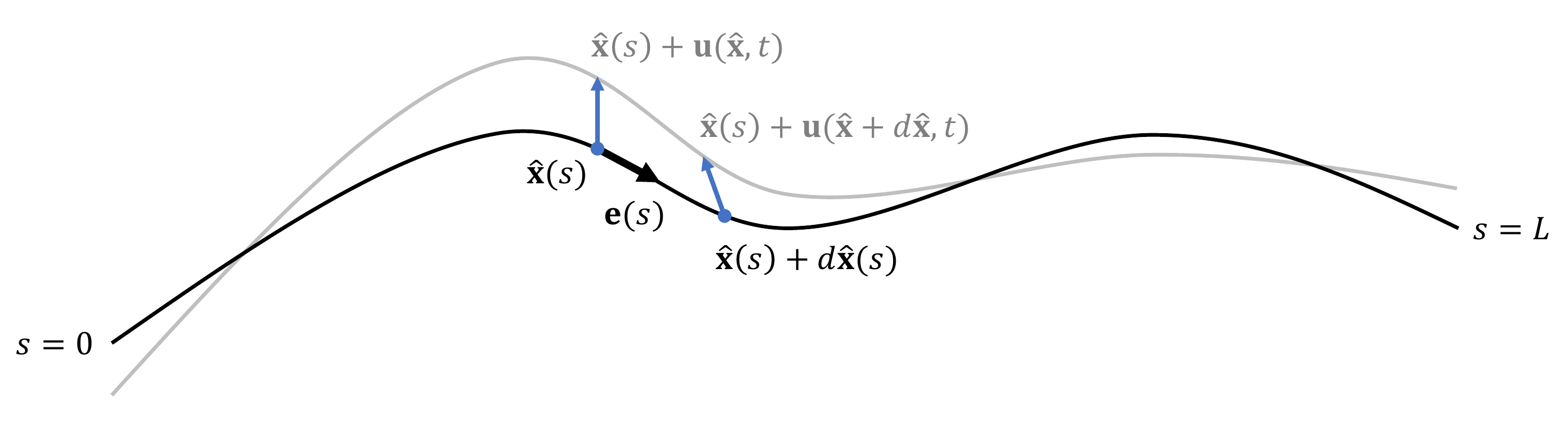}}
\caption{Schematic illustration of fibre deformation. The undeformed fibre, shown as black curve, is represented by the position vector $\hat{\mathbf{x}}(s)$, which is parameterised in terms of the arc length $s$. The cable starts at $s=0$ and ends at $s=L$. Under a displacement field $\mathbf{u}(\hat{\mathbf{x}},t)$, displayed as blue arrows, the Lagrangian position $\hat{\mathbf{x}}(s)$ along the undeformed fibre moves to $\hat{\mathbf{x}}(s)+d\hat{\mathbf{x}}(s)$. The result is the deformed fibre, shown in grey. The local tangent vector $\mathbf{e}(s)$ is shown as a thick black arrow.}\label{F:geometry}
\end{figure}
\end{center}

We begin with the derivation of an exact relation between the deformation tensor $\mathbf{F}(\mathbf{x},t)$ along the fibre and the traveltime $T(t)$ of a pulse that propagates from the beginning to the end of the fibre. The only assumption is that $T(t)$ is much smaller than the characteristic time scales of deformation, meaning that the fibre does not deform significantly while a pulse is propagating.\\[5pt]
To ease calculations, we adopt a parameterised representation of the fibre, with its position $\hat{\mathbf{x}}(s)$ given in terms of the arc length $s$. The latter ranges between $0$ and the total length of the fibre $L$, as shown in figure \ref{F:geometry}.

%
%

\subsection{Reference case of an undeformed fibre}

We begin with the (hypothetical) reference case where the fibre is not deformed. The time it takes for a pulse to travel from fibre location $\hat{\mathbf{x}}(s)$ to the neighbouring location $\hat{\mathbf{x}}(s)+d\hat{\mathbf{x}}(s)$ is given by
\begin{equation}\label{E:000}
dT = \frac{|d\hat{\mathbf{x}}(s)|}{c[\hat{\mathbf{x}}(s)]}\,,
\end{equation}
where $c[\hat{\mathbf{x}}(s)]$ is the potentially space-dependent speed of light along the fibre. The coordinate $\hat{\mathbf{x}}$ is interpreted as a Lagrangian coordinate, meaning that it co-moves with the deforming fibre instead of being attached to the static reference frame. By definition of the arc length, we can express the total traveltime of the pulse through the fibre as
\begin{equation}\label{E:002}
T = \int\limits_{s=0}^L \frac{ds}{c[\hat{\mathbf{x}}(s)]}\,.
\end{equation}

\subsection{Traveltimes under an arbitrary deformation field}

Under deformation, position $\hat{\mathbf{x}}$ moves to a new position $\hat{\mathbf{x}}+\mathbf{u}(\hat{\mathbf{x}},t)$, where $\mathbf{u}(\hat{\mathbf{x}},t)$ is the (seismic) displacement field, as illustrated in figure \ref{F:geometry}. To avoid clutter, we omit dependencies on $s$ for the moment. The neighbouring point at the original position $\hat{\mathbf{x}}+d\hat{\mathbf{x}}$ moves to $\hat{\mathbf{x}}+d\hat{\mathbf{x}}+\mathbf{u}(\hat{\mathbf{x}}+d\hat{\mathbf{x}},t)$. It follows that the traveltime of the pulse within the deformed segment of the cable is now
\begin{equation}\label{E:100}
dT(t) = \frac{|d\hat{\mathbf{x}}+\mathbf{u}(\hat{\mathbf{x}}+d\hat{\mathbf{x}},t)-\mathbf{u}(\hat{\mathbf{x}},t)|}{c[\hat{\mathbf{x}},\mathbf{u}(\hat{\mathbf{x}},t)]}\,.
\end{equation}
The denominator accounts for the fact that the speed of light may be a function of the deformation of the fibre. Since $d\hat{\mathbf{x}}$ is infinitesimally small, we can rewrite the numerator as
\begin{equation}\label{E:101}
\mathbf{u}(\hat{\mathbf{x}}+d\hat{\mathbf{x}},t)-\mathbf{u}(\hat{\mathbf{x}},t) = \mathbf{F}(\hat{\mathbf{x}},t)\, d\hat{\mathbf{x}}\,,
\end{equation}
where the Cartesian components of the deformation tensor $\mathbf{F}$ are defined by
\begin{equation}\label{E:102}
F_{ij} = \frac{\partial u_i}{\partial x_j}\,.
\end{equation}
In terms of the deformation tensor, we can rewrite (\ref{E:100}) as
\begin{equation}\label{E:103}
dT(t) = \frac{|d\hat{\mathbf{x}}+\mathbf{F}(\hat{\mathbf{x}},t)\, d\hat{\mathbf{x}}|}{c[\hat{\mathbf{x}},\mathbf{u}(\hat{\mathbf{x}},t)]}\,.
\end{equation}
This can be further simplified using the arc-length parameterisation of the position vector $\mathbf{x}$. In fact, we find
\begin{equation}\label{E:104}
d\hat{\mathbf{x}} = \frac{d\hat{\mathbf{x}}(s)}{ds}\,ds = \mathbf{e}(s)\, ds\,,
\end{equation}
where $\mathbf{e}(s)$ is the normalised tangent vector along the fibre. With this, we obtain
\begin{equation}\label{E:105}
dT(t) = \frac{|[\mathbf{I}+\mathbf{F}(\hat{\mathbf{x}},t)]\,\mathbf{e}(s)|}{c[\hat{\mathbf{x}},\mathbf{u}(\hat{\mathbf{x}},t)]} ds\,,
\end{equation}
and the total, time-dependent traveltime of the pulse becomes
\begin{equation}\label{E:106}
T(t) = \int\limits_{s=0}^L \frac{|[\mathbf{I}+\mathbf{F}(\hat{\mathbf{x}},t)]\,\mathbf{e}(s)|}{c[\hat{\mathbf{x}},\mathbf{u}(\hat{\mathbf{x}},t)]} ds\,.
\end{equation}

\subsection{Phase changes in monochromatic signals}

In the specific case of a monochromatic input with circular frequency $\omega$, the traveltime difference $\Delta T(t) = T(t) - T$ translates into a phase difference
\begin{equation}\label{E:200}
\phi(t) = \omega \Delta T(t)\,,
\end{equation}
between the reference and the deformed state. Substituting (\ref{E:002}) and (\ref{E:106}), we obtain
\begin{equation}\label{E:201}
\phi(t) = \omega \int\limits_{s=0}^L \frac{|[\mathbf{I}+\mathbf{F}(\hat{\mathbf{x}},t)]\,\mathbf{e}(s)|}{c[\hat{\mathbf{x}},\mathbf{u}(\hat{\mathbf{x}},t)]} ds - \omega \int\limits_{s=0}^L \frac{ds}{c[\hat{\mathbf{x}}(s)]}\,.
\end{equation}
Taking the time derivative $\partial_t$ of (\ref{E:201}), yields the phase changes with respect to time,
\begin{equation}\label{E:202}
\boxed{\partial_t\phi(t) = \omega \partial_t \int\limits_{s=0}^L \frac{|[\mathbf{I}+\mathbf{F}(\hat{\mathbf{x}},t)]\,\mathbf{e}(s)|}{c[\hat{\mathbf{x}},\mathbf{u}(\hat{\mathbf{x}},t)]} ds\,.}
\end{equation}
Equation (\ref{E:202}) is valid without any approximations, and it relates measured phase changes of the monochromatic laser signal to the deformation field $\mathbf{u}(\hat{\mathbf{x}},t)$ along the fibre.

\section{First-order approximations}\label{S:first}

While being exact, equation (\ref{E:202}) is often too complicated to be practically useful. It can be simplified considerably by realising that typical seismic displacement fields $\mathbf{u}$ have amplitudes in the nano- or micrometre range. Therefore, the norm of the deformation tensor $\mathbf{F}$ is typically six or more orders of magnitude smaller than $1$. It follows that first-order approximations can easily be justified in many applications. To avoid clumsy notation, we work with a slight reformulation of equation (\ref{E:202}), which uses the refractive index $r=c_0/c$, where $c_0$ is the speed of light in vacuum.

\subsection{Relation to the strain tensor and DAS measurements}

Equation (\ref{E:202}) can be simplified substantially under the assumption that deformation is small compared to $1$.  For this, we first note that
\begin{equation}\label{E:300a}
|[\mathbf{I}+\mathbf{F}]\,\mathbf{e}|^2 = \w{e}^T (\w{F}^T + \w{I}^T) (\w{F} + \w{I}) \w{e} = \w{e}^T \w{F}^T \w{F} \w{e} + \w{e}^T \w{F}^T \w{e} + \w{e}^T \w{F} \w{e} + \w{e}^T \w{e}\,.
\end{equation}
Neglecting the second-order term involving $\w{F}^T \w{F}$ and realising that $\w{e}^T \w{e}=1$ by definition of the unit tangent vector, we obtain
\begin{equation}\label{E:300}
|[\mathbf{I}+\mathbf{F}]\,\mathbf{e}|^2 \overset{.}{=} 1 + 2\mathbf{e}^T \mathbf{E} \mathbf{e}\,,
\end{equation}
with the strain tensor $\mathbf{E}=(\mathbf{F}^T + \mathbf{F})/2$. and $\overset{.}{=}$ meaning correct to first order in deformation quantities. Denoting the axial strain along the fibre as $\epsilon=\mathbf{e}^T \mathbf{E} \mathbf{e}$ and using the first-order relation $\sqrt{1+2\epsilon} \overset{.}{=} 1 + \epsilon$, we arrive at
\begin{equation}\label{E:301}
\partial_t\phi(t) \overset{.}{=} \frac{\omega}{c_0}\, \partial_t \int\limits_{s=0}^L r[\hat{\mathbf{x}},\mathbf{u}(\hat{\mathbf{x}},t)] \, \left( 1 + \epsilon[\hat{\mathbf{x}}(s),t] \right)\, ds\,.
\end{equation}
The dependence of the refractive index $r$ on the deformation of the fibre is called the photoelastic effect, and it is primarily a dependence on the axial strain $\epsilon$. Using the first-order Taylor approximation $r(\epsilon)\overset{.}{=}r_0 + r'\epsilon$, we obtain the relation
\begin{equation}\label{E:302}
\boxed{ \partial_t\phi(t) \overset{.}{=} \frac{\omega}{c_0}\, \partial_t \int\limits_{s=0}^L r_\text{eff}[\hat{\mathbf{x}}(s)] \, \epsilon[\hat{\mathbf{x}}(s),t]\, ds\,, }
\end{equation}
where $r_\text{eff}=r_0 + r'$ is the effective refractive index. It is defined as the sum of the static refractive index $r_0$ and the axial strain derivative $r'$, meaning that it takes the photoelastic effect into account.  The derivative $r'$ is commonly expressed in terms of the strain coefficient $\xi$ as $r' = r_0 (\xi-1)$, with an experimentally determined value of $\xi\approx 0.78$ \citep{Bertholds_1988}. Equation (\ref{E:302}) provides a direct relation between phase changes $\partial_t\phi$ measured by the transmission system, and the axial strain rate $\partial_t\epsilon$. In the case where the refractive index is roughly constant along the fibre, it suffices to integrate - or, for simplicity, sum - DAS measurements of $\partial_t\epsilon$ along the fibre in order to synthesise transmission measurements of $\partial_t\phi$.

\subsection{Illustrative and educational examples}

To better understand the nature and the consequences of the first-order approximation, we continue with a series of simple examples. While they may not be realised exactly in practice, they are still educationally valuable.

\subsubsection{Deformation states}

\begin{center}
\begin{figure}
\center\scalebox{0.48}{\includegraphics{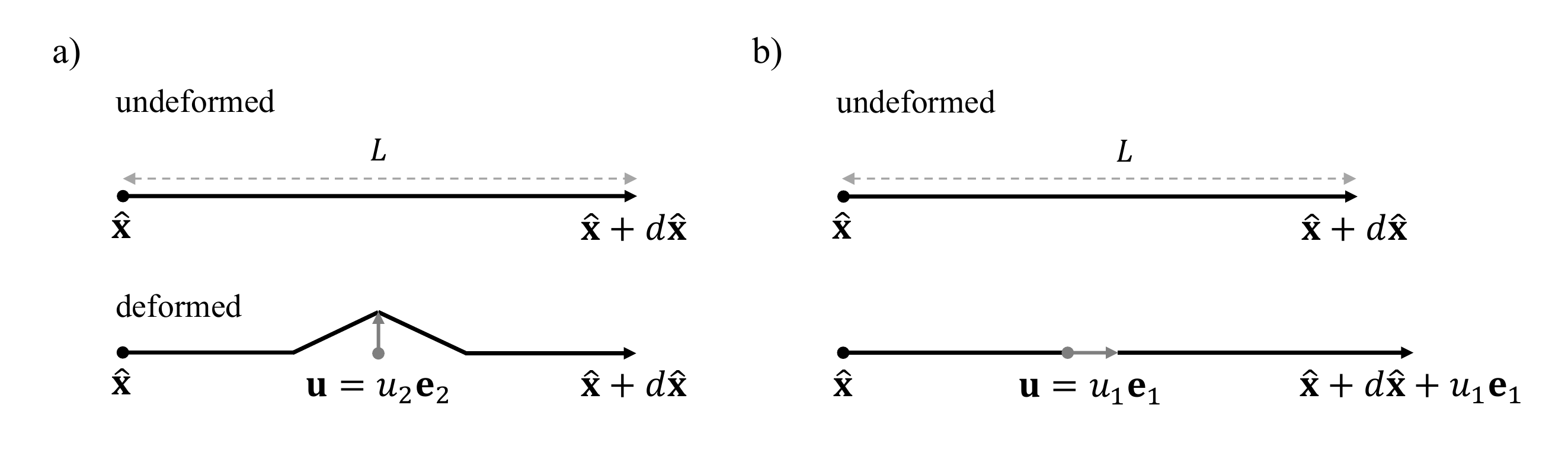}}
\caption{Simple examples of fibre deformation. a) Deformation of an originally straight fibre perpendicular to its orientation, i.e., by an amount $u_2$ in $\w{e}_2$ direction. b) Deformation of a straight fibre in the direction of its orientation, i.e., by an amount $u_1$ in $\w{e}_1$ direction.}\label{F:examples1}
\end{figure}
\end{center}

If a phase change is of first or second order in displacement $\mathbf{u}$ or deformation $\mathbf{F}$, depends on the geometry of fibre deformation. As a simple illustration, we consider the example in figure \ref{F:examples1}a, where the displacement $\mathbf{u}=u_2 \mathbf{e}_2$ is localised and perpendicular to the fibre direction $\mathbf{e}_1$. The length of the undeformed fibre is $L$, and the length of the deformed fibre is
\begin{equation}\label{E:410}
L' = 2 \sqrt{(L/2)^2 + u_2^2}\,.
\end{equation}
Expanding $L'$ into a Taylor series around $u_2=0$, yields
\begin{equation}\label{E:411}
L' = L + \frac{2}{L} u_y^2 + \mathcal{O}(u_y^3)\,.
\end{equation}
Equation (\ref{E:411}) implies that local deformation perpendicular to the fibre orientation is of second order in the displacement. Hence, there is no first-order effect on traveltimes and phase changes, and the second-order effect may be the only one observable, provided that $u_y$ is large enough to raise the observation above the noise level. \\[5pt]
An observable consequence of a dominant second-order effect is frequency doubling. To see this, we consider a harmonic deformation, $u_y = \sin(2\pi f t)$, where $f$ is frequency. Substituting $u_y$ into equation (\ref{E:411}), we find
\begin{equation}\label{E:412}
L'(t) = L + \frac{1}{L}\left[ 1 - \cos(4 \pi f t)   \right] + ...\,,
\end{equation}
It follows that the resulting length and phase changes oscillate with twice the frequency, $2f$, of the actual deformation.\\[5pt]
We contrast the above example with a deformation style where the displacement is parallel to the fibre orientation, $\mathbf{u}=u_1 \mathbf{e}_1$, as shown in figure \ref{F:examples1}b. Trivially, the length of the deformed fibre is
\begin{equation}\label{E:413}
L' = L + u_1\,.
\end{equation}
Hence, deformation along the fibre orientation has a first-order effect in $\mathbf{u}$ on length and phase changes. In the case of a harmonic deformation, the observed phase changes will have the same frequency as the deformation.

%
%

\subsubsection{Fibre geometry}

\begin{center}
\begin{figure}
\center\scalebox{0.48}{\includegraphics{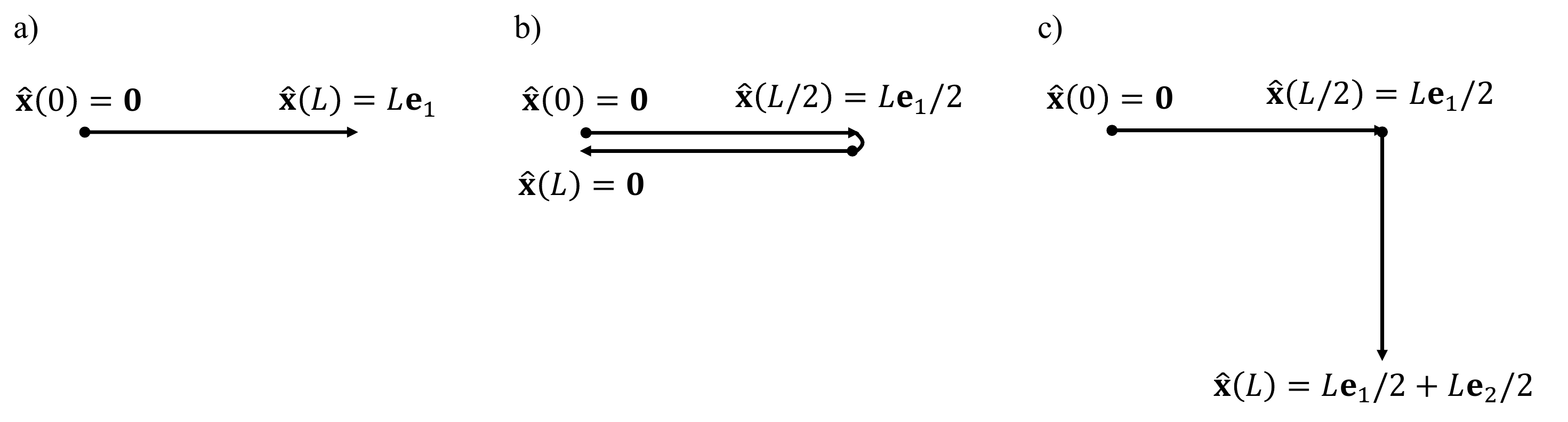}}
\caption{Simple examples of fibre geometries and corresponding arc length parameterisations.}\label{F:examples}
\end{figure}
\end{center}

When second-order effects and changes in the refractive index throughout the fibre can be ignored, we merely need to solve the integral
\begin{equation}\label{E:400}
I = \int\limits_{s=0}^L \epsilon[\hat{\mathbf{x}}(s),t]\,ds\,,
\end{equation}
in order to predict phase changes. To start simple, we let the fibre follow a straight line from $\hat{\mathbf{x}}(0)=\mathbf{0}$ to $\hat{\mathbf{x}}(L)=L\mathbf{e}_1$, where $\mathbf{e}_1$ is the unit vector in $x_1$-direction and therefore also the tangent vector $\mathbf{e}$. This fibre geometry is shown in figure \ref{F:examples}a. Evaluating (\ref{E:400}), gives
\begin{equation}\label{E:401}
I = \int\limits_{s=0}^L \frac{\partial}{\partial x_1} u_1(s\mathbf{e}_1) ds = \int\limits_{s=0}^L \frac{\partial}{\partial x_1} u_1(x_1\mathbf{e}_1) dx_1 = u_1(L\mathbf{e}_1) - u_1(\mathbf{0})\,.
\end{equation}
For a perfectly straight fibre, it follows that only the displacement at the beginning and the and of the fibre are measured, correct to first order. Hence, if the perturbation of interest is only present in between these two points, not affecting the beginning and end points, nothing can be measured. 
%
%
In the second example, we again consider a fibre parallel to the $\mathbf{e}_1$ direction, which, however, returns to its starting point, as shown in figure \ref{F:examples}b. The parameterisation of the fibre is now $\hat{\mathbf{x}}(s)=s\mathbf{e}_1$ for the first half, with $s$ between $0$ and $L/2$. For the second half, it is $\hat{\mathbf{x}}(s)=L\mathbf{e}_1 - s\mathbf{e}_1$, with $s$ between $L/2$ and $L$. Using this to evaluate the integral (\ref{E:400}), we find
\begin{equation}\label{E:402}
I = 2 u_1(L\mathbf{e}_1/2) - 2 u_1(\mathbf{0})\,.
\end{equation}
Hence, we again observe that only the start and the end point make a contribution.\\[5pt]
The fibre in our final example, shown in figure \ref{F:examples}c, features a $90^\circ$ kink half way between the start and end points. The parameterisation is $\hat{\mathbf{x}}(s)=s\mathbf{e}_1$ for $s$ between $0$ and $L/2$. For the second part it is $\hat{\mathbf{s}}=L\mathbf{e}_1/2 + (s-L/2)\mathbf{e}_2$. Again evaluating the integral (\ref{E:400}), yields
\begin{equation}\label{E:403}
I = u_1(L\mathbf{e}_1/2) - u_1(\mathbf{0}) + u_2(L\mathbf{e}_1/2+L\mathbf{e}_2/2) - u_2(L\mathbf{e}_1/2)\,.
\end{equation}
Interestingly, in addition to the end points $\hat{\mathbf{x}}(0)=\mathbf{0}$ and $\hat{\mathbf{x}}(s)=L\mathbf{e}_1/2+L\mathbf{e}_2/2$, also the kink point at $\hat{\mathbf{x}}(L/2)=L\mathbf{e}_1/2$ now makes a contribution. This suggests that curvature of the cable may play some role.

\subsection{The role of fibre curvature and directional fibre sensitivity}

To gain deeper understanding of these results, we go a few steps back and return to equation (\ref{E:100}). Denoting by $\hat{\mathbf{u}}(s,t) = \mathbf{u}[\hat{\mathbf{x}}(s),t]$ the displacement field along the fibre, we obtain
\begin{equation}\label{E:500}
|d\hat{\mathbf{x}}+\mathbf{u}(\hat{\mathbf{x}}+d\hat{\mathbf{x}},t)-\mathbf{u}(\hat{\mathbf{x}},t)| = |d\hat{\mathbf{x}}+\hat{\mathbf{u}}(s+ds,t)-\hat{\mathbf{u}}(s,t)| = \left| \mathbf{e}(s) + \partial_s \hat{\mathbf{u}}(s,t) \right| ds\,,
\end{equation} 
where $\partial_s=\partial/\partial s$ denotes the derivative with respect to the arc length $s$. Again making use of the first-order relation
\begin{equation}\label{E:501}
\left| \mathbf{e}(s) + \partial_s \hat{\mathbf{u}}(s,t) \right| = 1 + \mathbf{e}(s)^T \partial_s \hat{\mathbf{u}}(s,t)\,,
\end{equation}
and the Taylor expansion of the refractive index, yields a new version of equation (\ref{E:302})
\begin{equation}\label{E:502}
\partial_t\phi(t) \overset{.}{=} \frac{\omega}{c_0}\, \partial_t \int\limits_{s=0}^L \hat{r}_\text{eff}(s)\,\mathbf{e}(s)^T \partial_s\hat{\mathbf{u}}(s,t)\, ds\,,
\end{equation}
where we defined, for convenience, $\hat{r}_\text{eff}(s)=r_\text{eff}[\hat{\mathbf{x}}(s)]$. Using the product rule,
\begin{equation}\label{E:503}
\hat{r}_\text{eff}(s)\,\mathbf{e}(s)^T \partial_s\hat{\mathbf{u}}(s,t) = \partial_s \left[ \hat{r}_\text{eff}(s)\, \mathbf{e}(s)^T  \hat{\mathbf{u}}(s,t) \right] - \partial_s \left[ \hat{r}_\text{eff} (s) \, \mathbf{e}(s)^T \right] \, \hat{\mathbf{u}}(s,t)\,,
\end{equation}
and substituting back into equation (\ref{E:302}), we find
\begin{equation}\label{E:504}
\boxed{\partial_t\phi(t) \overset{.}{=} \underbrace{ \frac{\omega}{c_0}\, \partial_t \left. \left[ \hat{r}_\text{eff}(s) \, \mathbf{e}(s)^T \hat{\mathbf{u}}(s,t) \right] \right|_{s=0}^{s=L} }_{\text{start/end point contribution}}- \underbrace{ \frac{\omega}{c_0}\, \int\limits_{s=0}^L \partial_s \left[ \hat{r}_\text{eff}(s)\,\mathbf{e}(s)^T  \right] \, \partial_t\hat{\mathbf{u}}(s,t)\, ds}_{\text{curvature contribution}}\,.}
\end{equation}
Equation (\ref{E:504}) has two contributions to the observed phase changes. The first one originates from the displacement at the start and end points of the fibre. It vanishes when the start and end points are not affected by deformation. This may happen in cases of rather localised deformation that only happens along a smaller section of the fibre.\\[5pt]
The second contribution results from changes of the tangent vector and the refractive index along the fibre. In fact, the term
\begin{equation}\label{E:505}
\w{a}(s) = -\frac{\omega}{c_0}\, \partial_s \left[ \hat{r}_\text{eff}(s)\,\mathbf{e}(s)  \right]
\end{equation}
plays the role of a directional fibre sensitivity. Assuming that deformation does not significantly affect the end points, we may use the directional sensitivity to rewrite (\ref{E:504}) as
\begin{equation}\label{E:506}
\boxed{\partial_t\phi(t) \overset{.}{=}  \int\limits_{s=0}^L \w{a}(s)^T \, \partial_t\hat{\mathbf{u}}(s,t)\, ds\,.}
\end{equation}
The amplitude of the directional sensitivity is proportional to changes of the refractive index  along the fibre, and to changes of the tangent vector $\w{e}$. The latter is equivalent to the fibre being curved.

\section{Numerical examples and estimation of second-order effects}\label{S:Second}

\begin{center}
\begin{figure}
\center\scalebox{0.55}{\includegraphics{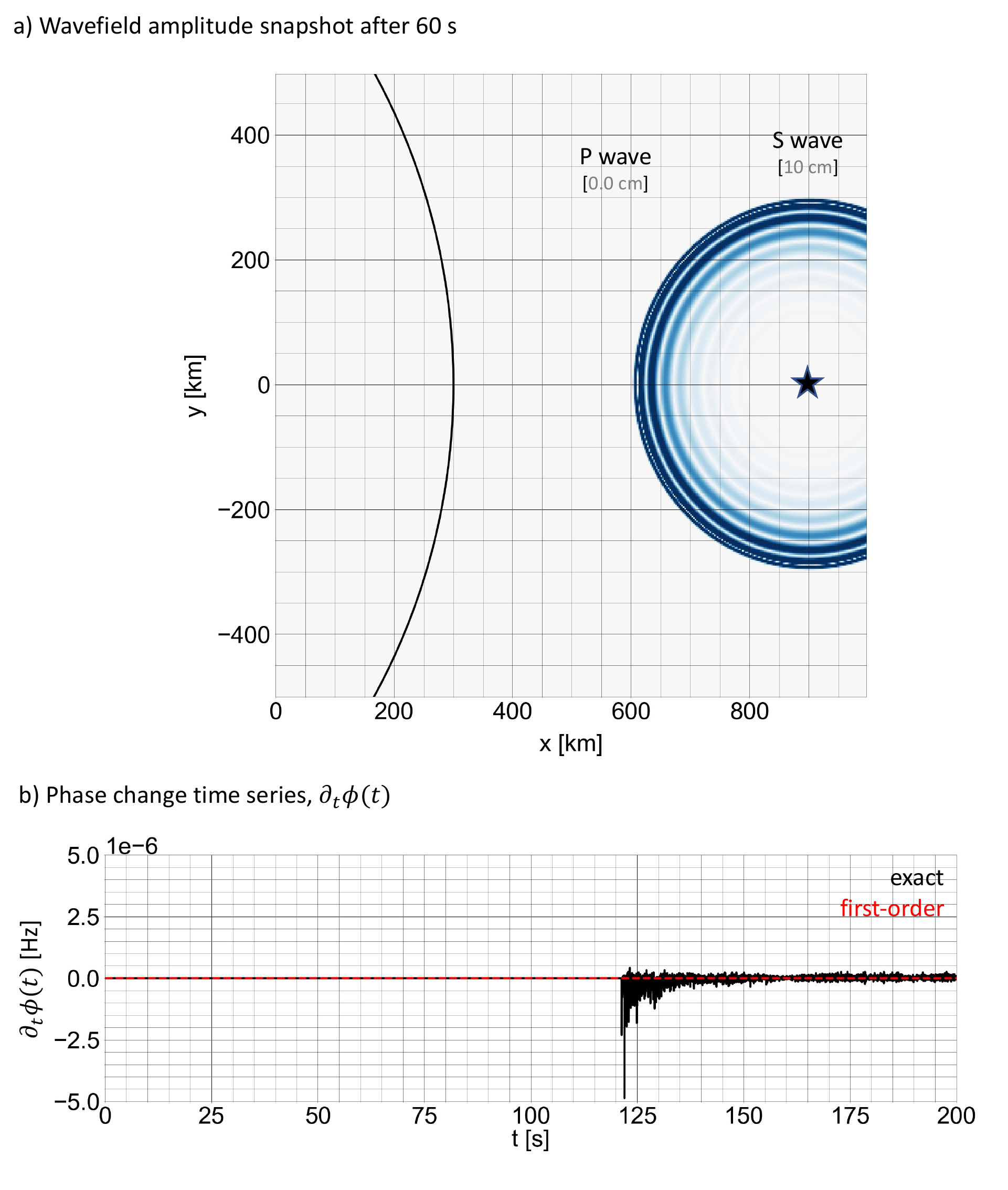}}
\caption{Exact solution vs. first-order approximation. a) Wavefield amplitude snapshot for a source that only produces an S wave within the $x$-$y$-plane where the fibre, shown as back curve, is located.  The fibre has the shape of a circular segement with non-zero curvature, in order to ensure that the cable sensitivity is non-zero. The P wave label with $0$ cm amplitude is near the position where the P wave front would be at that time, had it been excited. This pathological case is similar to the example in figure \ref{F:examples1}a. Note that the amplitude of the S wave ($10$ cm) is chosen to be unrealistically large in order to ensure that the second-order effect is not contaminated by floating-point errors. b) Phase change time series for the exact solution (black) and the first-order approximation (red).}\label{F:ex02}
\end{figure}
\end{center}

\begin{center}
\begin{figure}
\center\scalebox{0.55}{\includegraphics{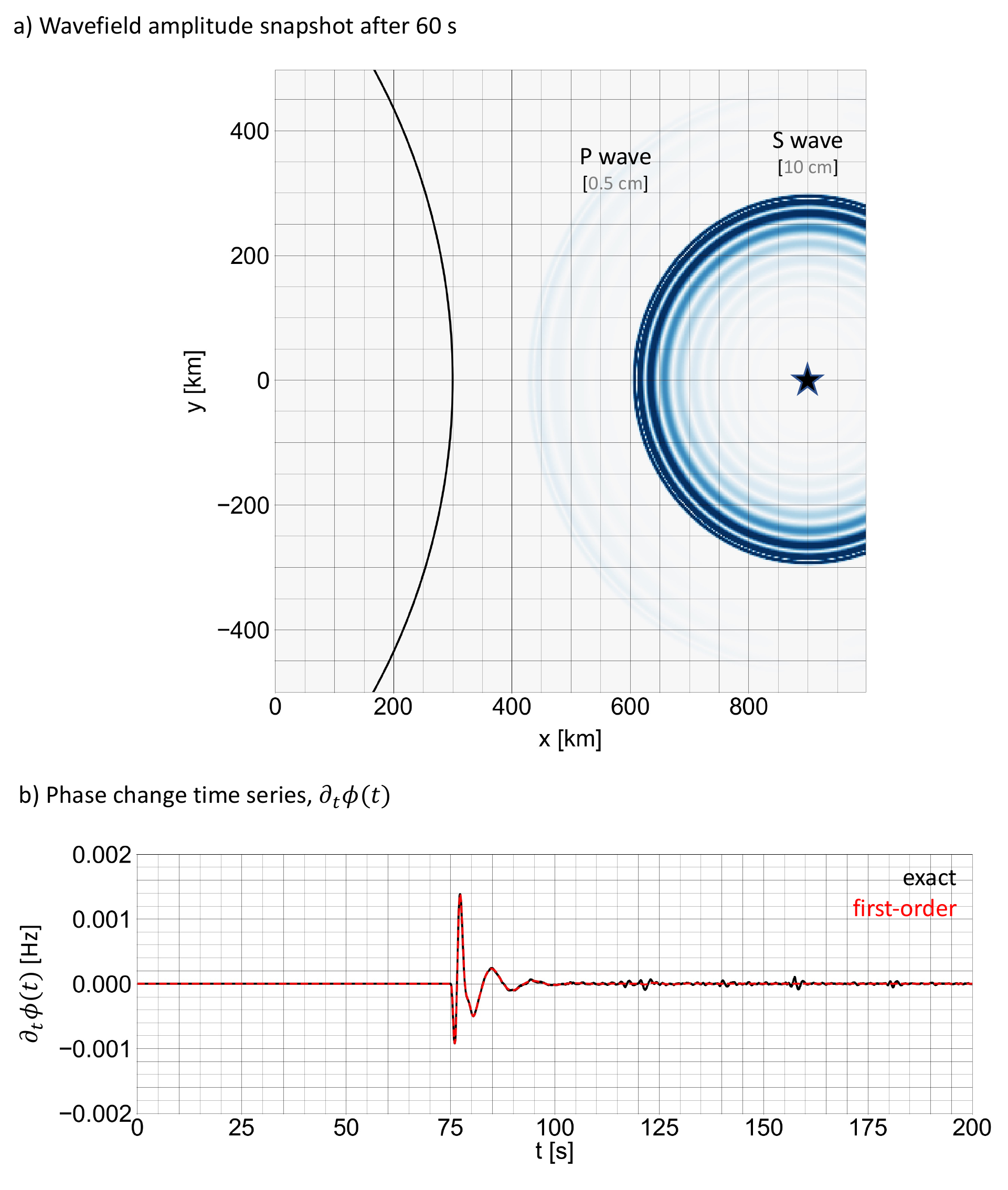}}
\caption{Exact solution vs. first-order approximation. a) Wavefield amplitude snapshot for a source that produces a small P wave ($0.5$ cm amplitude) and a large S wave ($10$ cm amplitude) within the $x$-$y$-plane where the fibre, shown as black curve, is located. Despite the unrealistically large amplitude of the S wave, which is responsible for the second-order effect, the first-order P wave contribution dominates the phase change time series in panel b).}\label{F:ex01}
\end{figure}
\end{center}

The following examples are intended to estimate the second-order contributions to the phase change signals $\partial_t\phi(t)$. We make the plausible assumption that the effective refractive index $\hat{r}_\text{eff}(s)$ is constant over a seismic wavelength, i.e., $\mathcal{O}(10)$ km in our examples, thereby allowing us to ignore its derivative. Not trying to mimic a specific acquisition system, we set $\omega \hat{r}_\text{eff} / c=1$ m$/$s$^2$ for simplicity. With this setting, we have
\begin{equation}\label{E:600}
\frac{\omega}{c_0}\, \partial_t \int\limits_{s=0}^L \partial_s \left( \hat{r}_\text{eff}(s)\,\mathbf{e}(s)^T  \right) \, \hat{\mathbf{u}}(s,t)\,dt =  \int\limits_{s=0}^L \w{n}(s)^T   \partial_t \hat{\mathbf{u}}(s,t)\,ds\,,
\end{equation}
where $\w{n}$ is the non-normalised normal vector $\frac{d}{ds}\w{e}(s)$. To avoid any complications, the elastic medium for our calculations is unbounded, isotropic and perfectly elastic, with P velocity $\alpha=8000$ m$/$s, S velocity $\beta=5000$ m$/$s, and density $\rho=3000$ kg$/$m$^3$. Well-known analytical solutions for moment tensor and single force sources may be found, e.g., in \citet{Aki_Richards_2002}.\\[5pt]
In our first example, summarised in figure \ref{F:ex02}, we consider a single force acting in the $z$-direction. The cable, having the geometry of a circular segment in the $x$-$y$-plane, therefore only records the motion of an S wave, with polarisation in $z$-direction. This scenario corresponds to the example from figure \ref{F:examples1}a, where we deformed a fibre in the direction perpendicular to its orientation. There, we found, as a consequence of Pythagoras' theorem, that this style of deformation does not produce a first-order effect on the transmitted phase. Therefore, the higher-order effects may be visible. In fact, this is what we observe in figure \ref{F:ex02}b. Using the exact, non-linear forward modelling equation (\ref{E:202}), we obtain a phase change time series that features an oscillation with frequencies much higher than the $0.1$ Hz maximum frequency of the incoming wave. The signal starts to be visible at the time when the S wave reaches the point of the cable that is closest to the source. The amplitude of the signal then decays rapidly but remains at high frequency. Using the first-order approximation from equation (\ref{E:504}) does not reproduce this effect. Instead, as expected in the absence of any first-order effects, it produces a time series that is identically zero for all times.\\[5pt]
The calculation of higher-order effects can be a numerical challenge because they can be so small that floating point errors start to be important. This is also the case here, mostly because of the term $|(\w{I}+\w{F})\,\w{e}|$, which requires us to compute the square root of a number that differs from $1$ only by a number that is usually many orders of magnitude smaller than $1$. To avoid this problem, the amplitude of the S wave in this example is unrealistically large, on the order of $10$ cm. Such large displacements can only be observed near the epicentres of large earthquakes. Displacements in the micrometre range, would have been a more sensible choice from a seismological perspective.
%
%
In our second example, we slightly change the orientation of the single force source, to have a small component in $x$-$y$-direction, such that it produces a P wave recorded by the fibre in the $x$-$y$-plane, as shown in figure \ref{F:ex01}a. Though the P wave has an amplitude that is around $20$ times smaller than the amplitude of the S wave, it does produce a first order effect that completely overwhelms the higher-order effects induced by the S wave, which are already unrealistically large. The agreement between the exact forward model (\ref{E:202}) and the linear approximation (\ref{E:504}) confirms the validity of the latter, even for displacement fields that are large compared to the majority of earthquake wavefields.
%
%

\section{Discussion and conclusions}\label{S:Discussion}

We have developed a theory for deformation sensing using measurements of phase changes in transmitted laser pulses. Though the exact relation between deformation and phase change measurements is non-linear, useful first-order approximations can be derived and justified for seismological applications where strain is typically much smaller than $1$. The first of these relations connects phase changes to an integral of axial strain along the fibre, thereby enabling a direct comparison to measurements of distributed strain, e.g., using DAS. The second relation establishes a link between the displacement field along the fibre and a directional fibre sensitivity. This sensitivity is proportional to spatial changes in refractive index and the curvature of the fibre.\\[5pt] 
In the following, we provide additional discussion on simplifications and their range of validity, and of higher-order effects.

\subsection{Simplifications}

The relation between observed optical phase changes $\partial_t\phi$ and deformation $\w{F}$ is already complicated by the geometry of the fibre and the potential relevance of non-linearity. While these mathematical complexities can be handled with reasonable effort and simplifications, there is also a range of practical issues that contribute to measurement errors in actual experiments, and that may be more challenging to control or quantify.\\[5pt]
Often, the geometry of the fibre is only known to within metres or tens of metres. Fibre-optic cables deployed on glaciers or unstable slopes, for instance, may move considerably during the experiment.  Telecommunication cables in particular may follow highly irregular paths that can only be estimated by coarse tap testing. The length and curvature of tightly wrapped cable segments, commonly used to provide some buffer that prevents tearing of the cable under long-term strain, may also not be well constrained, though they can make a significant contribution to the measurements, as shown in equation (\ref{E:504}).\\[5pt]
A convenient but potentially profound simplification in our developments is the assumption that deformation of the medium, for example, the Earth, equals deformation of the fibre. Such perfect coupling is unlikely to be realised in practice, where the layout of fibre-optic cables must conform to the boundary conditions of a potentially harsh field site. Telecommunication cables, evidently, are not deployed for seismic experiments. Therefore, coupling may be somewhat random, and good coupling is plainly a matter of luck.\\[5pt]
Finally, a whole family of optical effects are likely to make an addition to the measurement errors.  Optical fibres exhibit variations in the shape of their core along the fibre length, and they may experience non-uniform stress that breaks their cylindrical symmetry.  The result is quasi-random and frequency-dependent birefringence that leads to polarisation-mode dispersion (PMD) and an average differential group delay (DGD) $\Delta\tau$ between the two different polarisation axes.  For fibre lengths $\gg\!1$ km,  DGD can be expressed as $\Delta\tau = D\sqrt{\ell}$, where $D$ is the PMD coefficient and $\ell$ the length of the fibre segment \citep[e.g.][]{Agrawal_2012}.  For modern telecommunication fibres, $D$ should be below $0.5$ ps$/$km$^{1/2}$ \citep{ITU_2022}, thereby providing a quantitative measure of optical fibre heterogeneity that may be compared to the effect of fibre curvature on a case-by-case basis.  Experimental research suggests that PMD at least plays a negligible role compared to the photoelastic effect \citep{Butter_1978}.\\[5pt]
While it is certainly possible to reduce measurement errors by improving our knowledge of cable geometry, coupling, and optical properties, an equally important effort is the design of suitable measurement functionals that honour the nature of the observational errors. During the past three decades, numerous measurement functionals have been developed for seismic inversion based on conventional seismometer recordings \citep[e.g.,][]{Luo_Schuster_1991,Gee_1992,Fichtner_et_al_2008,Brossier_2009b,vanLeeuwen_2010,Bozdag_2011,Rickers_2012}.
This effort may now need to be repeated for fibre-optic seismology.

\subsection{Higher-order effects}

As illustrated by equation (\ref{E:202}), a fibre-optic phase transmission system exhibits a non-linear relation between the displacement field $\w{u}$ and the observed output signal $\partial_t\phi$. The higher-order effects are characterised by higher frequencies, and they may be observable in cases where first-order effects are absent.\\[5pt] 
This may be achieved in laboratory settings where the mode of deformation can be simple and controlled. However, in seismological applications it seems unlikely that first-order effects can be avoided, mainly for two reasons: (1) Realistic fibre geometries are not perfect, meaning that some fibre segments are likely to be oriented such that the wavefield produces a first-order effect. (2) The seismic wavefield is complex, due to the 3-D heterogeneity and the presence of the Earth's surface with topography. Asymptotically, the seismic wavefield equipartitions, meaning that wave states with all possible propagation directions and polarisations will eventually occur, thereby producing some first-order effect.\\[5pt]
The positive consequence is that first-order effects will usually dominate the phase change observations by far. This means that linear forward modelling equations can be used, thereby facilitating the comparison to DAS data, as well as the calculation of sensitivity kernels with respect to Earth structure or source parameters.


\begin{acknowledgments}
This work was partially funded by the Real-time Earthquake Risk Reduction for a Resilient Europe project (RISE) under the European Union’s Horizon 2020 research and innovation programme (grant agreement No 821115).  Open-source code in the form of Python Jupyter Notebooks is available at \href{https://github.com/afichtner/TransmissionFibreOptics}{https://github.com/afichtner/TransmissionFibreOptics}.
\end{acknowledgments}


\bibliography{biblio}
 
\end{document}